\documentclass[conference]{IEEEtran}
\IEEEoverridecommandlockouts
\usepackage{cite}
\usepackage{multirow}
\usepackage{amsmath,amssymb,amsfonts}
\usepackage{algorithmic}
\usepackage{graphicx}
\usepackage{subcaption}
\usepackage{xcolor}
\usepackage{comment}
\usepackage{caption}

\def\BibTeX{{\rm B\kern-.05em{\sc i\kern-.025em b}\kern-.08em
    T\kern-.1667em\lower.7ex\hbox{E}\kern-.125emX}}
\begin{document}

\title{{\huge On the Bit Error Performance of OTFS Modulation using Discrete Zak Transform}}
\author{\IEEEauthorblockN{Vineetha Yogesh, Vighnesh S Bhat, Sandesh Rao Mattu, and A. Chockalingam }
\IEEEauthorblockA{Department of ECE, Indian Institute of Science, Bangalore 560012 }}
\maketitle

\begin{abstract}
In orthogonal time frequency space (OTFS) modulation, Zak transform approach is a natural approach for converting information symbols multiplexed in the DD domain directly to time domain for transmission, and vice versa at the receiver. Past research on OTFS has primarily considered a two-step approach where DD domain symbols are first converted to time-frequency domain which are then converted to time domain for transmission, and vice versa at the receiver. The Zak transform approach can offer performance and complexity benefits compared to the two-step approach. This paper presents an early investigation on the bit error performance of OTFS realized using discrete Zak transform (DZT). We develop a compact DD domain input-output relation for DZT-OTFS using matrix decomposition that is valid for both integer and fractional delay-Dopplers. We analyze the bit error performance of DZT-OTFS using pairwise error probability analysis and simulations. Simulation results show that 1) both DZT-OTFS and two-step OTFS perform better than OFDM, and 2) DZT-OTFS achieves better performance compared to two-step OTFS over a wide range of Doppler spreads. 
\end{abstract}
\vspace{1mm}

\section{Introduction}
\label{sec1}
As we rapidly progress towards the next-generation wireless technologies, i.e., 6G and beyond, the air interface and the modulation waveform must cater to diverse scenarios and use cases, which include high-Doppler wireless environments that result in rapidly time-varying channels. Orthogonal time frequency space (OTFS) modulation has gained attention due to its ability to enable highly reliable communication over high-Doppler channels \cite{b1}-\cite{b13}. OTFS modulation utilizes the delay-Doppler (DD) domain to mount information-carrying symbols. Also, it represents the time-varying channel in the DD domain, which is almost time-invariant and sparse.

In wireless communication based signal processing applications, Zak transform \cite{zak1} is emerging as an important tool for directly transforming a signal in time domain to delay-Doppler (DD) domain and vice versa. In the context of OTFS modulation, Zak transform approach provides a natural means to directly convert the information symbols multiplexed in the DD domain to time domain for transmission, and vice versa at the receiver. Whereas past research on OTFS has mainly considered a two-step approach for DD domain to time domain conversion and vice versa. In the two-step approach, information symbols in the DD domain are converted to time-frequency (TF) domain using inverse symplectic finite Fourier transform (ISFFT) which are then converted to time domain for transmission using Heisenberg transform. Corresponding inverse transforms are carried out at the receiver to convert the received time domain signal to DD domain in two steps. 

While the two-step approach has been inspired by its compatibility with existing multicarrier modulation waveforms (e.g., OFDM) and a majority of the existing work on OTFS consider this approach, the Zak approach remains less explored so far. Consequently, this paper focuses on the Zak approach and its performance in comparison with that of the two-step approach. The Zak approach in continuous time has been derived from first principles in \cite{b14}, where it is shown that the spectral efficiency performance of the Zak approach remains invariant to user velocity, whereas it degrades in OFDM as the velocity increases. However, \cite{b14} does not provide a discrete-time system model for the Zak approach and the bit error rate (BER) performance. The work in \cite{b16} considers the Zak approach only at the receiver while using a two-step approach at the  transmitter, and presents the BER performance. Recently, \cite{b15} has presented a discrete Zak transform (DZT) approach both at the transmitter and receiver, but it does not provide the BER performance. Also, it does not provide a compact input-output relation in a matrix-vector form amenable for BER performance evaluation. In the above context, the new contributions in this paper can be summarized as follows.
\begin{itemize}
\item We first develop a compact DD domain input-output relation in matrix-vector form for DZT-OTFS (with Zak approach at the transmitter as well as the receiver) using matrix decomposition that is valid for both integer and fractional delay-Dopplers. This compact vectorized input-output relation is an useful contribution as it enables the BER performance evaluation (through analysis and simulations) and the development of techniques and algorithms for DZT-OTFS transceivers. 
\item Using the developed vectorized input-output relation, we analyze the asymptotic diversity of DZT-OTFS. We also present the simulated BER performance of DZT-OTFS in comparison with that of the two-step approach of OTFS as well as OFDM over a range of Doppler spreads.
\item Our simulation results show that $i$) both DZT-OTFS and two-step OTFS perform better than OFDM, and $ii$) DZT-OTFS achieves better performance compared to two-step OTFS over a wide range of Doppler spreads. 
\end{itemize}

\begin{figure*}[t]
\includegraphics[width=\linewidth]{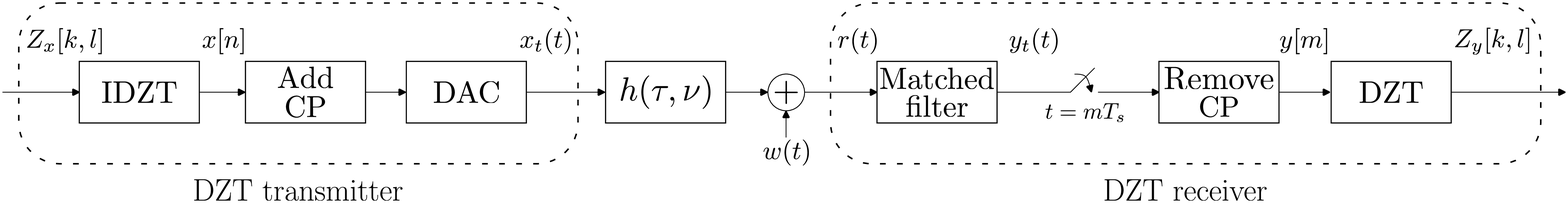}
\caption{Block diagram of DZT-OTFS transceiver.}
\label{OTFS_block_diagram}
\end{figure*}
The rest of the paper is organized as follows. The DZT-OTFS system model and its vectorized DD domain input-output relation are developed in Sec. \ref{sec2}. The diversity analysis of DZT-OTFS is presented in Sec. \ref{sec3}. BER simulation results and discussions are presented in Sec. \ref{sec4}. Conclusions are presented in Sec. \ref{sec5}.

\textit{Notations:} Matrices and vectors are denoted by upper and lower case boldface letters, respectively, $\text{diag}\{v_i\}$ denotes a diagonal matrix with diagonal entries $v_i, i=1,2,\cdots$, and $\mathbf{I}_K$ denotes the identity matrix of size $K$. $(\cdot)_N$ denotes the modulo $N$ operation. Hermitian, transpose, and conjugation operations are denoted by $(\cdot)^H$, $(\cdot)^T$, and $(\cdot)^*$, respectively. Hadamard product, kronecker product, and convolution operations are denoted by $\odot$, $\otimes$ and $\circledast$, respectively. 

\section{DZT-OTFS system model}
\label{sec2}
In this section, we present the DZT-OTFS system model and develop the compact DD domain input-output relation in matrix-vector form. Figure \ref{OTFS_block_diagram} shows the block diagram of a DZT-OTFS transceiver.
OTFS is a two-dimensional modulation technique that uses a DD grid to multiplex information symbols from a modulation alphabet ${\mathbb A}$. 
In OTFS using DZT, $KL$ information symbols, denoted by $Z_x[k,l]$s, are multiplexed over a $K \times L$ DD grid given by $\Big\{ \left(k\Delta \nu=\tfrac{k}{KLT_s}, l \Delta \tau =lT_s\right), k=0,\cdots,K-1,  l=0,\cdots, L-1\Big\}$, where $T_s=1/B$ is the symbol duration, $B$ is the bandwidth available for communication, and $\Delta \tau$ and $\Delta \nu$ are the delay and Doppler resolutions, respectively. The DD domain symbols, $Z_x[k,l]$s, are converted to TD using inverse DZT (IDZT) \cite{b15}, as
\begin{align}
x[n] = \frac{1}{\sqrt{K}}\sum_{k=0}^{K-1}Z_{x}[k,n \text{ mod } L]e^{j2\pi \frac{\left\lfloor n/L \right\rfloor}{K}k}.
\label{IDZT_eqn}
\end{align} 
To mitigate the inter-frame interference, a cyclic prefix (CP) of length $N_{CP} \geq \left\lceil{\tau_{\max}/T_{s}}\right\rceil$ is added in the TD, where $\tau_{\max}$ is the maximum delay spread of the channel and $\left\lceil{\cdot}\right\rceil$ denotes the ceil function. The TD sequence is then mounted on the time-shifted transmit pulse $g_{tx}(t), 0 \leq t \leq T_{s}$, resulting in a continuous time signal $x_{t}(t)$, given by 
\begin{align}
x_{t}(t) = \sum_{n=0}^{N+N_{CP}-1}x[(n-N_{CP})_N] g_{tx}(t-nT_s),
\label{s_t_eqn}
\end{align}
where $0\leq t \leq (N+N_{CP})T_{s}$ and $N=KL$.
This operation is carried out in the digital-to-analog converter (DAC) block in Fig. \ref{OTFS_block_diagram}. The transmitted signal $x_{t}(t)$ passes through a time-varying channel whose DD domain response is given by \cite{b4}
\begin{align}
h(\tau,\nu) = \sum_{i=1}^{P} h_{i} \delta(\tau - \tau_i) \delta(\nu - \nu_i),
\label{h_tau_nu}
\end{align}
where $P$ denotes the number of resolvable paths in the DD domain, $h_{i}$ is the channel gain of the $i$th path, and $\tau_i = (\alpha_{i}+a_{i}) T_{s}$ is the delay associated with the $i$th path, where $\alpha_{i}$ is integer part and $a_{i}\in [-0.5, 0.5]$ is the fractional part of the $i$th path's delay. Similarly, $\nu_i = \frac{(\beta_{i}+b_{i})}{KLT_{s}}$ is the Doppler associated with the $i$th path, where  $\beta_{i}$ is integer part and $b_{i} \in [-0.5, 0.5]$ is the fractional part of the $i$th path's Doppler. The received time domain signal is given by 
\begin{align}
r(t)=\int_{\nu}\int_{\tau}h(\tau,\nu)x_{t}(t-\tau)e^{j2\pi\nu(t-\tau)}d\tau d\nu+w(t).
\label{r_t_eqn}
\end{align}
Substituting \eqref{s_t_eqn} and \eqref{h_tau_nu} in \eqref{r_t_eqn}, we have 
\begin{eqnarray}
r(t) & = &  \sum_{i=1}^{P} h_{i}\sum_{n=0}^{N+N_{CP}-1}x[(n-N_{CP})] \nonumber \\
& \hspace{-0mm} & \hspace{-0mm} g_{tx}(t-\tau_{i}-nT_s) e^{j2\pi\nu_{i}t}+ w(t),
\label{r_t_eqn_2}
\end{eqnarray}
where $w(t)$ is the additive noise. At the receiver, after matched filtering with receive pulse $g_{rx}(t)$, we get
\begin{align}
y_{t}(t)=\int_{\tau} r(\tau) g_{rx}^*(t-\tau) d\tau.
\label{y_t_t_eqn}
\end{align}
Let $g_{tx}(t) = g_{rx}(t) = p(t)$. Substituting \eqref{r_t_eqn_2} in \eqref{y_t_t_eqn}, we get   
\begin{multline}\label{y_t_t_eqn_2}
    y_{t}(t) =\sum_{i=1}^{P} h_{i}\hspace{-3mm}\sum_{n=0}^{N+N_{CP}-1} \hspace{-3mm}x[(n-N_{CP})_N] \\ \int_{\tau}p(\tau-\tau_{i}-nT_{s})p^*(t-\tau) e^{j2\pi\nu_{i}\tau}+v(t),
\end{multline}
where $v(t)$ is the match filtered noise. Assuming the pulse bandwidth to be much larger than the maximum Doppler, the integral in 
\eqref{y_t_t_eqn_2} can be approximated as \cite{b15} 
\begin{align}
     \int_{\tau}p(\tau-\tau_{i}-nT_{s})p^*(t-\tau) e^{j2\pi\nu_{i}\tau} d\tau \approx \nonumber\\
     e^{j2\pi\nu_{i}(\tau_{i}+nT_{s})} g(t-\tau_{i}-nT_{s}),
    \label{eqn_10}
\end{align}
where $g(t)  \triangleq \int_{\tau}p(\tau)p^*(t-\tau)d\tau$. For example, $g(t)$ for a raised cosine (RC) pulse with a roll-off factor $\gamma$ is given by
\begin{align}
    g(t)=\frac{\sin(\pi t/T_s)\cos(\gamma\pi t/T_s)}{(\pi t/T_s)(1-(2 \gamma t/T_s)^2)}.
    \label{pulse}
\end{align}
The output of the matched filter is sampled at $t = mT_{s}$, $m = 0, 1 , 2,\cdots$. The first $N_{CP}$ samples corresponding to the CP are discarded. The resulting discrete time-domain signal is given by 
\begin{align}
    y[m] \hspace{-0.5mm} = \hspace{-1.5mm} \sum_{i=1}^{P} h_{i}\hspace{-1mm}\sum_{n=0}^{N-1} \hspace{-1mm}x[n] g[(m-n)T_{s}-\tau_{i}]e^{j2\pi\nu_{i}(\tau_{i} \hspace{-0.5mm} + \hspace{-0.5mm} nT_{s})}+v[m].
    \label{eqn_11}
\end{align}
Substituting for $\tau_i$ and $\nu_i$ in (\ref{eqn_11}), we get
\begin{align}
    y[m] = \sum_{i=1}^{P} h_{i}e^{j2\pi\tau_i\nu_i}\hspace{-1mm}\sum_{n=0}^{N-1} \hspace{-1mm}x[n] e_{i}[n] g_{i}[m-n]+v[m],
    \label{eqn_12}
\end{align}
where $g_{i}[n] =g[n-l_{i}]$ and $e_{i}[n] =e^{j2\pi\frac{k_{i}}{KL}n}$, $n= 0,\cdots, N-1$, $l_i = \alpha_i+a_i$ and $k_i = \beta_i+b_i$. Here, $g_i$ and $e_i$ are the discrete sequences that capture the effect of delay and Doppler of the $i$th path in the channel, respectively. Let $h_{i}^{'} = h_{i}e^{j2\pi\tau_i\nu_i}$ and 
\begin{align}
y_{i}[m] =\sum^{N-1}_{n=0} x[n] e_{i}[n] g_{i}[m-n].
\label{io_eqn}
\end{align}
Then (\ref{eqn_12}) can be written as 
\begin{align}
    y[m] = \sum_{i=1}^{P} h_{i}^{'} y_{i}[m] +v[m].
    \label{eqn_15}
\end{align}

\subsection{Vectorized formulation of DD domain input-output relation}
\label{sec2a}
Here, we first write the TD input-output relation in (\ref{eqn_15}) in a vectorized form and transform it to a DD domain input-output relation in vectorized form using matrix decomposition as follows. The relation between $y_i[m], x[m], e_i[m]$, and $g_i[m]$ can be expressed in vector form as
\begin{align}
    {\bf y}_{i} &= ({\bf x} \odot {\bf e}_{i}) \circledast {\bf g}_{i},
    \label{eqn_16}
\end{align}
where $\mathbf{y}_i,\mathbf{e}_i,\mathbf{g}_i \in \mathbb{C}^{1\times N}$ are obtained by stacking the elements $y_i[m], e_i[m]$ and $g_i[m]$, respectively. The input-output relation in (\ref{eqn_15}) can then be written in a vectorized form as
\begin{align}
    \textbf y &= \begin{bmatrix}
     1 & 1 & \cdots & 1  
  \end{bmatrix} \text{diag}\{h'_1,\cdots,h'_P\}\begin{bmatrix}
     \mathbf{y}_{1}\\ 
     \vdots \\
     \mathbf{y}_{p}
  \end{bmatrix}+\textbf v.
  \label{eqn_17}
\end{align}
The discrete time-domain signal $\mathbf{y}$ at the receiver is converted to the DD domain using DZT as
\begin{align}
Z_{y}[k,l] =\frac{1}{\sqrt{K}} \sum_{n=0}^{K-1}y[l+nL]e^{-j2\pi \frac{k}{K}n}.
\label{DZT_eqn}
\end{align}
The equivalent DD domain representations of \eqref{io_eqn} and \eqref{eqn_15} are given by
\begin{align}
\hspace{-2mm}
    Z_{y_{i}}[k,l] \hspace{-0.5mm} = \hspace{-1mm} \sum_{m=0}^{L-1} \sum_{n=0}^{K-1} Z_{x}[n,m]Z_{e_{i}}[k-n,m] Z_{g_{i}}[k,l-m],
    \label{eqn_18}
\end{align}
and
\begin{align}
    Z_{y}[k,l] = \sum_{i=1}^{P} h_{i}^{'} Z_{y_{i}}[k,l] + Z_{v}[k,l],
    \label{eqn_19}
\end{align}
respectively.

\begin{table*}[]
\centering
\begin{tabular}{|cccc|cccc|c|}
\hline
\multicolumn{4}{|c|}{\multirow{2}{*}{System Parameters}}                                                                                                                                                                                                                                                                                                    & \multicolumn{4}{c|}{No. of difference matrices with rank}                              & \multirow{2}{*}{Minimum rank} \\ \cline{5-8}
\multicolumn{4}{|c|}{}                                                                                                                                                                                                                                                                                                                                      & \multicolumn{1}{c|}{1}  & \multicolumn{1}{c|}{2}   & \multicolumn{1}{c|}{3}    & 4     &                               \\ \hline
\multicolumn{1}{|c|}{\multirow{4}{*}{\begin{tabular}[c]{@{}c@{}}System-1\\ $K=2$, $L=2$,\\ $P=2$\end{tabular}}} & \multicolumn{1}{c|}{\multirow{2}{*}{DD Profile-A}} & \multicolumn{1}{c|}{\multirow{2}{*}{\begin{tabular}[c]{@{}c@{}}$\mathbf{\tau}=[0,0]\Delta \tau$\\ $\mathbf{\nu}=[0,1]\Delta \nu$\end{tabular}}}                           & w/o PR   & \multicolumn{1}{c|}{32} & \multicolumn{1}{c|}{208} & \multicolumn{1}{c|}{-}    & -     & 1                             \\ \cline{4-9} 
\multicolumn{1}{|c|}{}                                                                                          & \multicolumn{1}{c|}{}                              & \multicolumn{1}{c|}{}                                                                                                                                                     & with  PR & \multicolumn{1}{c|}{0}  & \multicolumn{1}{c|}{240} & \multicolumn{1}{c|}{-}    & -     & 2                             \\ \cline{2-9} 
\multicolumn{1}{|c|}{}                                                                                          & \multicolumn{1}{c|}{\multirow{2}{*}{DD Profile-B}} & \multicolumn{1}{c|}{\multirow{2}{*}{\begin{tabular}[c]{@{}c@{}}$\mathbf{\tau}=[0,1]\Delta \tau$\\ $\mathbf{\nu}=[0,1]\Delta \nu$\end{tabular}}}                           & w/o PR   & \multicolumn{1}{c|}{0}  & \multicolumn{1}{c|}{240} & \multicolumn{1}{c|}{-}    & -     & 2                             \\ \cline{4-9} 
\multicolumn{1}{|c|}{}                                                                                          & \multicolumn{1}{c|}{}                              & \multicolumn{1}{c|}{}                                                                                                                                                     & with PR  & \multicolumn{1}{c|}{0}  & \multicolumn{1}{c|}{240} & \multicolumn{1}{c|}{-}    & -     & 2                             \\ \hline
\multicolumn{1}{|c|}{\multirow{4}{*}{\begin{tabular}[c]{@{}c@{}}System-2\\ $K=2$, $L=4$,\\ $P=4$\end{tabular}}} & \multicolumn{1}{c|}{\multirow{2}{*}{DD Profile-C}} & \multicolumn{1}{c|}{\multirow{2}{*}{\begin{tabular}[c]{@{}c@{}}$\mathbf{\tau}=[0,0,1,1]\Delta \tau$\\ $\mathbf{\nu}=[0,1,0,1]\Delta \nu$\end{tabular}}}                   & w/o PR   & \multicolumn{1}{c|}{0}  & \multicolumn{1}{c|}{136} & \multicolumn{1}{c|}{2072} & 63072 & 2                             \\ \cline{4-9} 
\multicolumn{1}{|c|}{}                                                                                          & \multicolumn{1}{c|}{}                              & \multicolumn{1}{c|}{}                                                                                                                                                     & with PR  & \multicolumn{1}{c|}{0}  & \multicolumn{1}{c|}{0}   & \multicolumn{1}{c|}{0}    & 65280 & 4                             \\ \cline{2-9} 
\multicolumn{1}{|c|}{}                                                                                          & \multicolumn{1}{c|}{\multirow{2}{*}{DD Profile-D}} & \multicolumn{1}{c|}{\multirow{2}{*}{\begin{tabular}[c]{@{}c@{}}$\mathbf{\tau}=[0.2,1.4,2.3,3.6]\Delta \tau$\\ $\mathbf{\nu}=[-0.3,1.4,0.8,-0.2]\Delta \nu$\end{tabular}}} & w/o PR   & \multicolumn{1}{c|}{0}  & \multicolumn{1}{c|}{0}   & \multicolumn{1}{c|}{0}    & 65280 & 4                             \\ \cline{4-9} 
\multicolumn{1}{|c|}{}                                                                                          & \multicolumn{1}{c|}{}                              & \multicolumn{1}{c|}{}                                                                                                                                                     & with PR  & \multicolumn{1}{c|}{0}  & \multicolumn{1}{c|}{0}   & \multicolumn{1}{c|}{0}    & 65280 & 4                             \\ \hline
\end{tabular}
\caption{Rank profile of the difference matrices.}
  \label{rank}
\vspace{-2mm}
\end{table*}

Now, let $\mathbf{Z}_{y_i}, \mathbf{Z}_{e_i}$ and $\mathbf{Z}_{g_i}$ denote the DZT of the sequences $\mathbf{y}_i, \mathbf{e}_i$ and $\mathbf{g}_i$, respectively. Let $\mathbf{B}_i$ be a $K \times K$ matrix whose $j$th row is 
\begin{align}
\mathbf{B}_i(j,:)=\mathbf{Z}^T_{e_i}(:,u)\mathbf{P}_K^{j-1}, 
\label{Bi}
\end{align}
where $u=0,\cdots,L-1$, $j=0,\cdots, K-1$, and $\mathbf{P}_K$ is a $K \times K$ permutation matrix. Further, define
\begin{align}
\mathbf{E}_i = \text{diag}\{\mathbf{B}_0,\cdots,\mathbf{B}_{L-1}\}.
\label{Ei}
\end{align}
Likewise, define a block matrix $\mathbf{A}=[\mathbf{A}_0,\cdots,\mathbf{A}_{L-1}]$, where $\mathbf{A}_u=\text{diag}\{\mathbf{Z}_{g_i}(:,u)\}$. Let $\mathbf{Q}_u$ be an $N \times N$ matrix, given by
\begin{align}
\mathbf{Q}_u=\mathbf{P}_L^{u-1} \otimes \mathbf{I}_{K},
\label{Q}
\end{align}
where $\mathbf{P}_L$ is an $L \times L$ permutation matrix. Further, define
\begin{align}
\mathbf{G}_i = \begin{bmatrix}
     \mathbf{AQ}_{0}\\ 
     \vdots \\
     \mathbf{AQ}_{L-1}
  \end{bmatrix}.
  \label{Gi}
  \end{align}
  Using $\mathbf{E}_i$ and $\mathbf{G}_i$, the effective channel matrix ${\bf H}$ is given by
  \begin{align}
   \mathbf{H} = \sum_{i=1}^{P} h_{i}^{'}\mathbf{E}_{i} \mathbf{G}_{i}.
\label{ch}
\end{align}

Vectorizing (\ref{eqn_19}) using the above, we obtain end-to-end input-output relation in DD domain as 
\begin{align}
    \mathbf{y}_{DD} = \mathbf{x}_{DD} \mathbf{H} + \mathbf{v}_{DD},  
\label{ydd}
\end{align}
where $\mathbf{y}_{DD},\mathbf{x}_{DD}, \mathbf{v}_{DD} \in \mathbb{C}^{1\times N}$, such that $\mathbf{y}_{DD}(k+Kl)=Z_y[k,l]$, $\mathbf{x}_{DD}(k+Kl)=Z_x[k,l]$, and $\mathbf{v}_{DD}(k+Kl)=Z_v[k,l]$, for $k=0,1,\cdots,K-1$ and $l=0,1,\cdots, L-1$, and $\mathbf{H} \in \mathbb{C}^{N\times N}$ is the end-to-end effective channel matrix in the DD domain. We use the compact matrix-vector representation of DZT-OTFS in (\ref{ydd}) for the diversity analysis and BER simulations in the subsequent sections. 

\section{Diversity analysis of DZT-OTFS}
\label{sec3}
For the purpose of diversity analysis, the input-output relation in (\ref{ydd}) can be written in an alternate form, as
\begin{align}
    \mathbf{y}_{DD} = \mathbf{h}\mathbf{X} +\mathbf{v}_{DD},  
\label{ydda}
\end{align}
where $\mathbf{X}$ is a $P \times N$
matrix defined as
\begin{align}
     \mathbf{X} = (\mathbf{I}_P \otimes \mathbf{x}_{DD}) \begin{bmatrix}
     \mathbf{G}_{1}\mathbf{E}_{1}\\ 
     \vdots \\
     \mathbf{G}_{P}\mathbf{E}_{P}
  \end{bmatrix},  
\label{x}
\end{align}
and $\mathbf{h}=[h'_1,\cdots, h'_P]$ is a $1 \times P$ vector.
Let $\mathbf{x}^i_{DD}$ and $\mathbf{x}^j_{DD}$ be two distinct DD symbol vectors as defined in (\ref{ydd}), and let the corresponding symbol matrices be $\mathbf{X}_i$ and $\mathbf{X}_j$. Let $\rho$ be the normalized signal-to-noise ratio (SNR) given by $\rho=1/N_0$, where $N_0$ is the noise power. Assuming maximum likelihood (ML) detection and perfect channel knowledge at the receiver, the conditional pairwise error probability (PEP) between $\mathbf{X}_i$ and $\mathbf{X}_j$ is given by
\begin{eqnarray}
P(\mathbf{x}^i_{DD}\rightarrow \mathbf{x}^j_{DD}|\mathbf{h}) & \hspace{-2mm}=& \hspace{-2mm} P(\left \| \mathbf{y-hX}_j \right \|^2 < \left \| \mathbf{y-hX}_i\right \|^2) \nonumber \\
& \hspace{-6mm} & \hspace{-6mm}= \text{Q}\left ( \sqrt {\frac{\rho \left \| \mathbf{h}(\mathbf{X}_i-\mathbf{X}_j) \right \|^2}{2}} \right ).
\label{PEP1}
\end{eqnarray}
Using Chernoff bound and averaging over the statistics of $\mathbf{h}$, the average PEP can be bounded as
\begin{align}
\hspace{-3mm}
    P(\mathbf{x}^i_{DD}\rightarrow \mathbf{x}^j_{DD})\leq \mathbb{E}_{\mathbf{h}}\left \{ \text{exp}\left ( \frac{\rho \left \| \mathbf{h}(\mathbf{X}_i-\mathbf{X}_j) \right \|^2}{4} \right ) \right \},
    \label{PEP2}
\end{align}
where $\mathbb{E}_\mathbf{h}$ denotes expectation operation.
Assuming the elements of $\mathbf{h}$ are independent and identically distributed complex Gaussian random variables with zero mean and variance $1/P$, \eqref{PEP2} can be simplified as  \cite{DavidTse}
\begin{align}
    P(\mathbf{x}^i_{DD}\rightarrow \mathbf{x}^j_{DD})\leq \prod_{l=1}^{r}\frac{1}{1+\frac{\rho \lambda_{lij}}{4P}},
    \label{PEP3}
\end{align}
where $r$ is the rank of the difference matrix $(\mathbf{X}_i-\mathbf{X}_j)$, and ${\lambda_{lij}}$ is the eigen value of $(\mathbf{X}_i-\mathbf{X}_j)(\mathbf{X}_i-\mathbf{X}_j)^H$. At high SNRs, using the approximation $1+\frac{\rho \lambda_{lij}}{4P} \approx \frac{\rho \lambda_{lij}}{4P}$, we can write
\begin{align}
    P(\mathbf{x}^i_{DD}\rightarrow \mathbf{x}^j_{DD})\leq  \frac{1}{\rho^r \prod_{l=1}^{r}\frac{\lambda_{lij}}{4P} }.
    \label{PEP4}
\end{align}
Using the above average PEP expression, an upper bound on the bit error probability, $P_e$, can be obtained as 
\begin{eqnarray}
P_e & \hspace{-2mm}\leq & \hspace{-2mm} \frac{1}{QKL \log_2 |\mathbb{A}|}\hspace{-1mm}\sum_{i=1}^{Q}\hspace{-1mm}\sum_{j=1, j\neq i}^{Q} \hspace{-4mm}d(\mathbf{x}^i_{DD},\mathbf{x}^j_{DD})P(\mathbf{x}^i_{DD}\rightarrow \mathbf{x}^j_{DD}) \nonumber \\
& \hspace{-6mm} & \hspace{-6mm}= \frac{1}{QKL \log_2 |\mathbb{A}|}\sum_{i=1}^{Q}\sum_{j=1, j\neq i}^{Q}\frac{d(\mathbf{x}^i_{DD},\mathbf{x}^j_{DD})\rho^{-r}}{ \prod_{l=1}^{r}\frac{\lambda_{lij}}{4P}},
\label{PEP5}
\end{eqnarray}
where $d(\mathbf{x}^i_{DD},\mathbf{x}^j_{DD})$ is the Hamming distance between $\mathbf{x}^i_{DD}$ and $\mathbf{x}^j_{DD}$, $Q=|\mathbb{A}^{KL}|$, and $|\cdot |$ denotes the cardinality of a set. Also, a lower bound on $P_e$ is obtained by summing the PEPs corresponding to the pairs $\mathbf{X}_i$ and $\mathbf{X}_j$ such that the rank of $(\mathbf{X}_i-\mathbf{X}_j)$ is least amongst all the pairs. From (\ref{PEP5}), it is observed that the exponent of the SNR is $r$ which is the rank of the symbol difference matrix $(\mathbf{X}_i-\mathbf{X}_j)$ and for all $i,j$, $i \neq j$ and the minimum rank dominates the overall BER. Therefore the diversity order (DO) is given by
\begin{align}
    \text{DO} = \min_{i,j \hspace{2mm} i\neq j} \hspace{2mm} \text{rank}(\mathbf{X}_i-\mathbf{X}_j).
    \label{PEP6}
\end{align}
We have observed through simulations that the rank profile of the difference matrices can be such that $r$ is less than $P$ for several $({\bf X}_i,{\bf X}_j)$ pairs. This means the full DD diversity of $P$ is not always achieved. To alleviate this, phase rotation (PR) of the transmit vector $\mathbf{x}_{DD}$ can be carried out before transmission. That is, instead of transmitting the $\mathbf{x}_{DD}$ vector as such, a phase rotated vector $\tilde{\mathbf{x}}_{DD}= \mathbf{\Theta} \hspace{0.25mm} \mathbf{x}_{DD}$ is transmitted, where $\mathbf{\Theta}$ is a PR matrix given by $\mathbf{\Theta} = \text{diag}\{e^{j\frac{q}{N}} \}$, $q = 0,\cdots, N-1$. Our simulations have shown that with this PR scheme, the full diversity of $P$ is achieved. The above observations are illustrated in Table \ref{rank} which shows the rank profile of difference matrices in two systems, namely System-1 with $K=L=P=2$ and System-2 with $K=2,L=P=4$. The rank profiles for these two systems with and without PR are shown. It can be seen that, without PR, 32 out of 240 rank matrices in System-1 with DD Profile-A have rank 1 (which is less than $P=2$), resulting in a diversity order of 1. Likewise, without PR, many difference matrices in System-2 with DD Profile-C have ranks 2 and 3 (which are less than $P=4$). On the other hand, with PR, all the ranks are $P$ in both the systems for all the DD Profiles-A to D, achieving the full diversity order of $P$. In the next section, we have verified these minimum rank based diversity orders through BER simulations.

\section{Results and discussions}
\label{sec4}
In this section, we first present the simulated BER results that validate the analytical BER bounds and diversity orders obtained in the previous section. We then present the simulated BER performance of DZT-OTFS for different system settings
in comparison with those of two-step OTFS and OFDM.

\subsection{Diversity performance of DZT-OTFS}
\begin{figure}
\includegraphics[width=9.25cm,height=6.25cm]{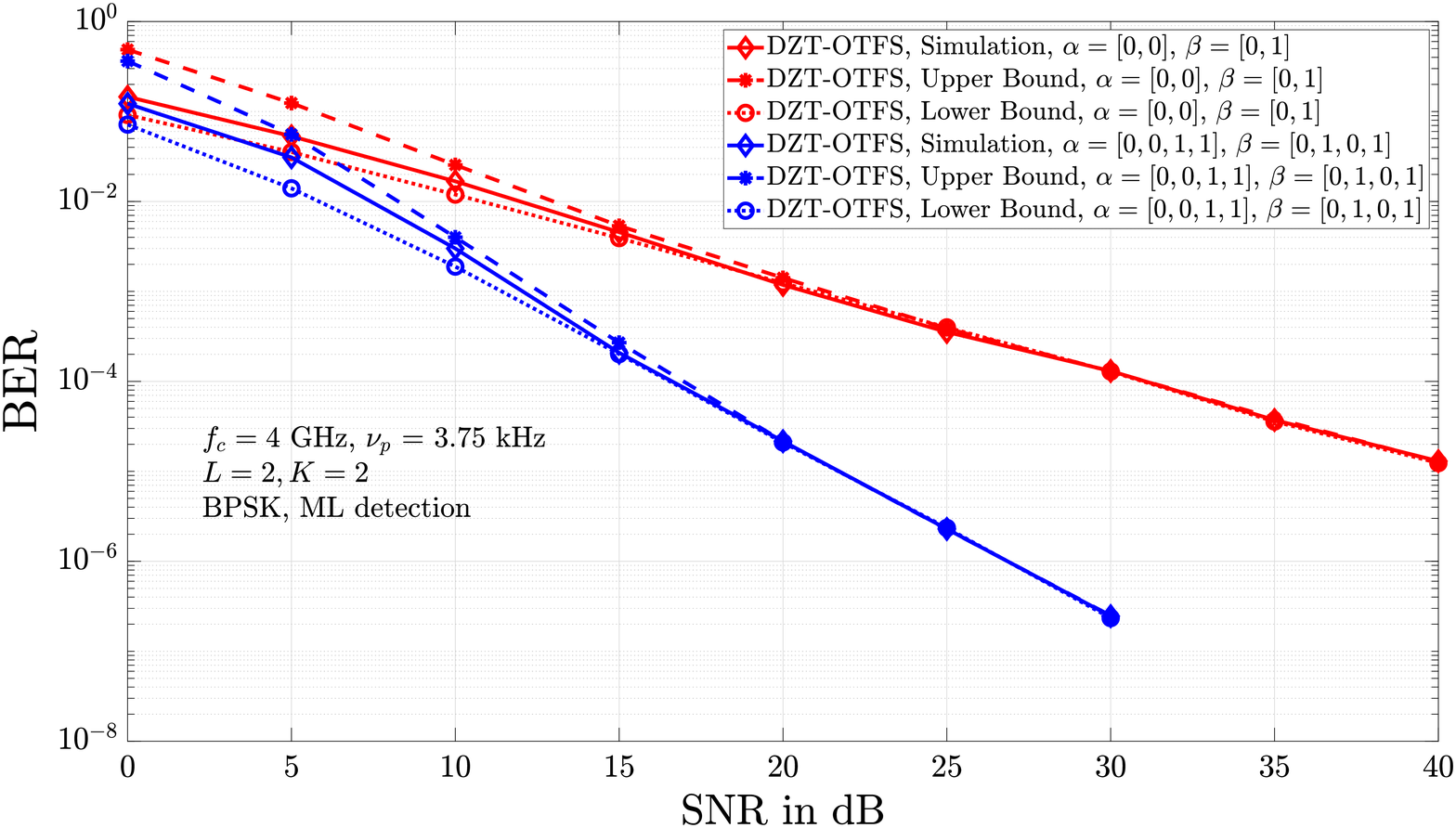}
\vspace{-4mm}
\caption{Simulated BER upper and lower bounds on BER for DZT-OTFS.}
\vspace{-4mm}
\label{fig2}
\end{figure}

Figure \ref{fig2} shows the simulated BER performance of a DZT-OTFS system with $K=L=2$ and $P=2,4$. A carrier frequency ($f_c$) of 4 GHz and $\nu_p$ of 3.75 kHz (hence, $\tau_p=\frac{1}{\nu_p}=0.267$ ms) are considered. The information symbols are chosen from BPSK modulation alphabet and ML detection is used at the receiver. Simulated BER curves without PR for System-1 with DD Profile-A and DD Profile-C in Table \ref{rank} are plotted. The corresponding upper bound and lower bound on BER are also plotted for comparison. 
The following observations can be made from the figure. The upper and lower bounds on BER and the simulated BER curves almost merge at high SNRs validating the analytical BER results. Also, the diversity slopes in the high SNR regime are 1 and 2, respectively, for System-1 with DD Profile-A and System-1 with DD Profile-C. These are in agreement with the diversity orders predicted according to minimum ranks shown in Table \ref{rank}.
\begin{figure}
\includegraphics[width=9.25cm,height=6.25cm]{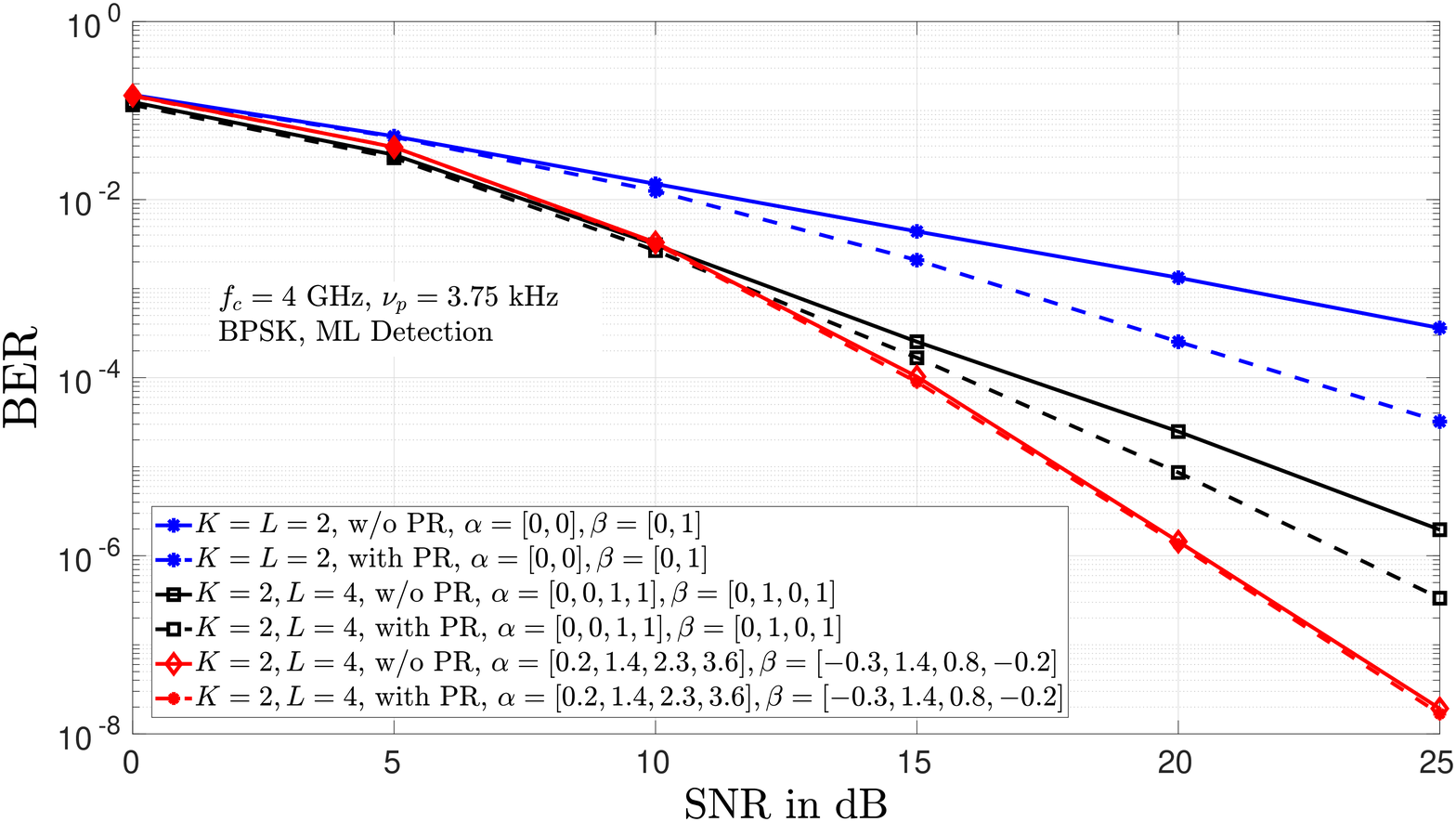}
\vspace{-4mm}
\caption{BER performance of DZT-OTFS with and without PR.}
\vspace{-4mm}
\label{fig3andfig4}
\end{figure}
\begin{figure}[t]
\includegraphics[width=9.25cm,height=6.25cm]{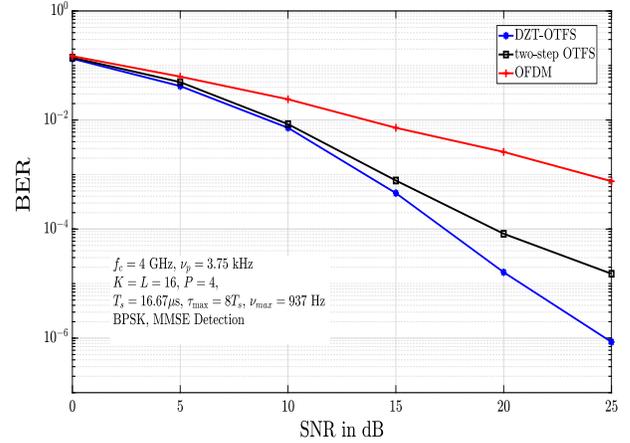}
\vspace{-4mm}
\caption{BER performance of DZT-OTFS, two-step OTFS, and OFDM as a function of SNR.}
\vspace{-4mm}
\label{fig5}
\end{figure}
\begin{figure}
\includegraphics[width=9.25cm,height=6.25cm]{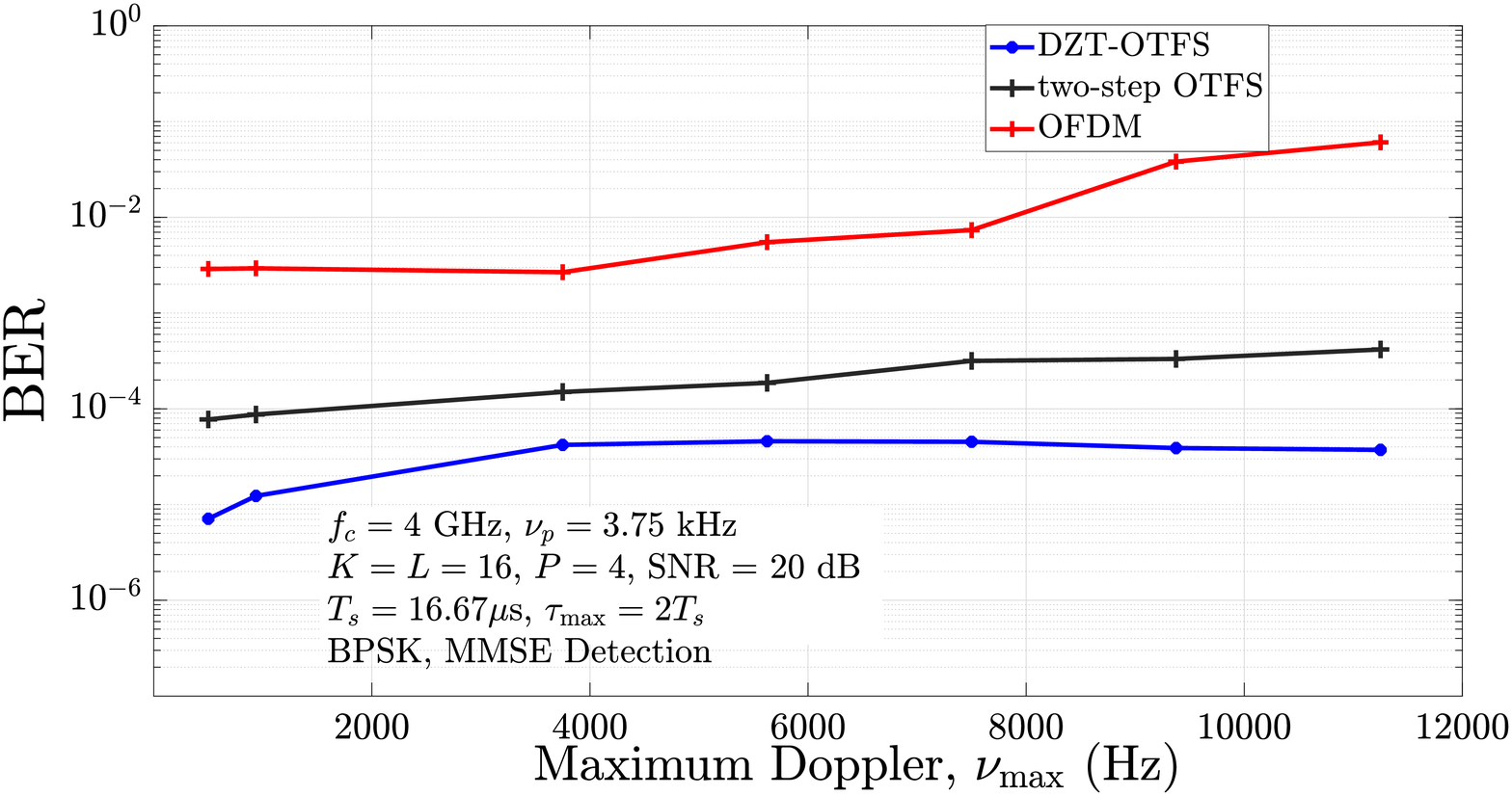}
\vspace{-4mm}
\caption{BER performance of DZT-OTFS, two-step OTFS, and OFDM as a function of maximum Doppler, $\nu_{\max}$.}
\vspace{-4mm}
\label{fig6}
\end{figure}

Next, Fig. \ref{fig3andfig4} shows the simulated BER curves without PR for 1) System-1 with DD Profile-A, 2) System-2 with DD Profile-C, and 3) System-2 with DD Profile-D. The corresponding BER curves with PR are also plotted. It is observed that the diversity slopes of the BER curves are in accordance with the minimum ranks shown in Table \ref{rank}. It is also observed that full diversity of $P$ is achieved with PR.

\subsection{BER performance of DZT-OTFS} 
\label{secc}
Here, we present the BER performance of DZT-OTFS for different system settings. The number of delay bins $L$ and the number of Doppler bins $K$ are chosen 
to meet the system specifications and channel constraints as follows. 
A fundamental rectangle in the DD domain of width $\tau_p$ and height $\nu_p$ is chosen such that $\tau_p \nu_p = 1$, and for a given available bandwidth ($B$) and delay requirement ($T$), $K$ and $L$ are chosen such that $L=\left\lceil{\frac{B}{\nu_p}}\right\rceil$ and $K=\left\lceil{\frac{T}{\tau_p}}\right\rceil$. The following parameters are considered in the simulations:
$B = 60$ kHz, $T = 4.27$ ms, and $T_s =\frac{1}{B}= 16.67\mu$s.
Fixing $\nu_p = 3.75$ kHz gives $\tau_p = \frac{1}{\nu_p}=0.267$ ms, $L=\left\lceil{\frac{B}{\nu_p}}\right\rceil=16$, and $K=\left\lceil{\frac{T}{\tau_p}}\right\rceil=16$. Fractional delay-Dopplers are considered in all the simulations and $g(t)$ is taken to be a RC pulse with roll-off factor $\gamma$. The channel is considered to have $P=4$ paths with uniform power delay profile. For a given maximum Doppler $\nu_{\max}$ and maximum delay $\tau_{\max}$, the $i$th path's $\alpha_i$ is a random integer uniformly sampled from $\{0,1,\cdots,\alpha_{\max}\}$, $a_i \in \mathrm{Unif}[-0.5,0.5]$, where $\alpha_{\max} = \text{round}\left({\frac{\tau_{\max}}{T_s}}\right)$, $\mathrm{Unif}[.,.]$ denotes uniform distribution, and $\nu_i=\nu_{\max}\cos(\theta)$, where $\theta \in \mathrm{Unif}[-\pi,\pi]$. The performance of two-step OTFS and OFDM are also obtained for comparison. BPSK modulation and minimum mean square error (MMSE) detection are used in all the systems.

\subsubsection{BER as a function of SNR}
Figure \ref{fig5} shows the BER performance of DZT-OTFS, two-step OTFS, and OFDM as a function of SNR. The maximum delay and Doppler spreads are taken to be $\tau_{\max} = 8Ts=133.33$ $\mu$s and $\nu_{\max} = 937$ Hz, and the roll-off factor of the RC pulse $\gamma$ is taken to be 0. From Fig. \ref{fig5}, it is observed that both DZT-OTFS and two-step OTFS achieve significantly better performance compared to OFDM. 
Also, DZT-OTFS is seen to achieve better performance compared to two-step OTFS, illustrating the inherent strength of the DZT-OTFS waveform compared to the two-step OTFS waveform. For example, at a BER of $10^{-3}$, DZT OTFS has an SNR gain of about 1 dB and 10 dB compared to two-step OTFS and OFDM, respectively. Further, the performance gap between DZT-OTFS and two-step OTFS increases in favor of DZT-OTFS as SNR increases.  

\begin{figure}
\includegraphics[width=9.25cm,height=6.25cm]{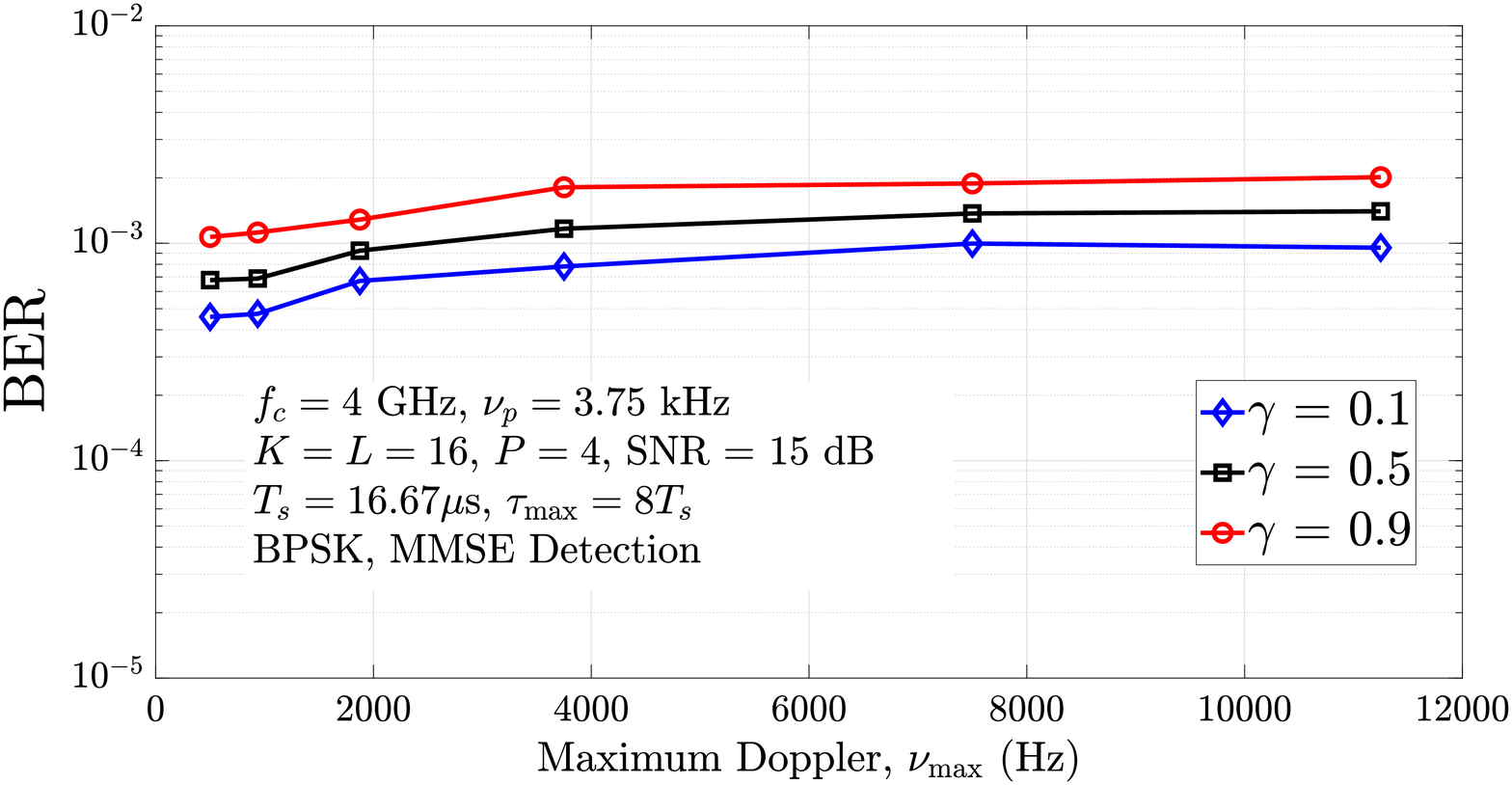}
\vspace{-4mm}
\caption{Effect of RC pulse roll-off factor, $\gamma$, on BER vs $\nu_{\max}$ performance.}
\vspace{-4mm}
\label{DZT_BER_vs_Doppler}    
\end{figure}

\subsubsection{BER as function of maximum Doppler, $\nu_{\max}$ } 
Figure \ref{fig6} shows the BER as a function of $\nu_{\max}$ at an SNR of 20 dB and $\gamma=0$. The simulations are done for a fixed $\tau_{\max} = 2T_s = 33.34$ $\mu$s and $\nu_{\max}$ varied in the range 500 Hz to 11.25 kHz. It is observed that, compared to OFDM, both DZT-OTFS and two-step OTFS are significantly more resilient to increase in maximum Doppler. In addition, DZT-OTFS is found to achieve better performance compared to two-step OTFS over a wide range of $\nu_{\max}$. This illustrates that DZT-OTFS waveform is more resilient to high Doppler spreads compared to two-step OTFS waveform.

\subsection{Effect of RC pulse roll-off factor}
\begin{figure}
\includegraphics[width=9.25cm,height=6.25cm]{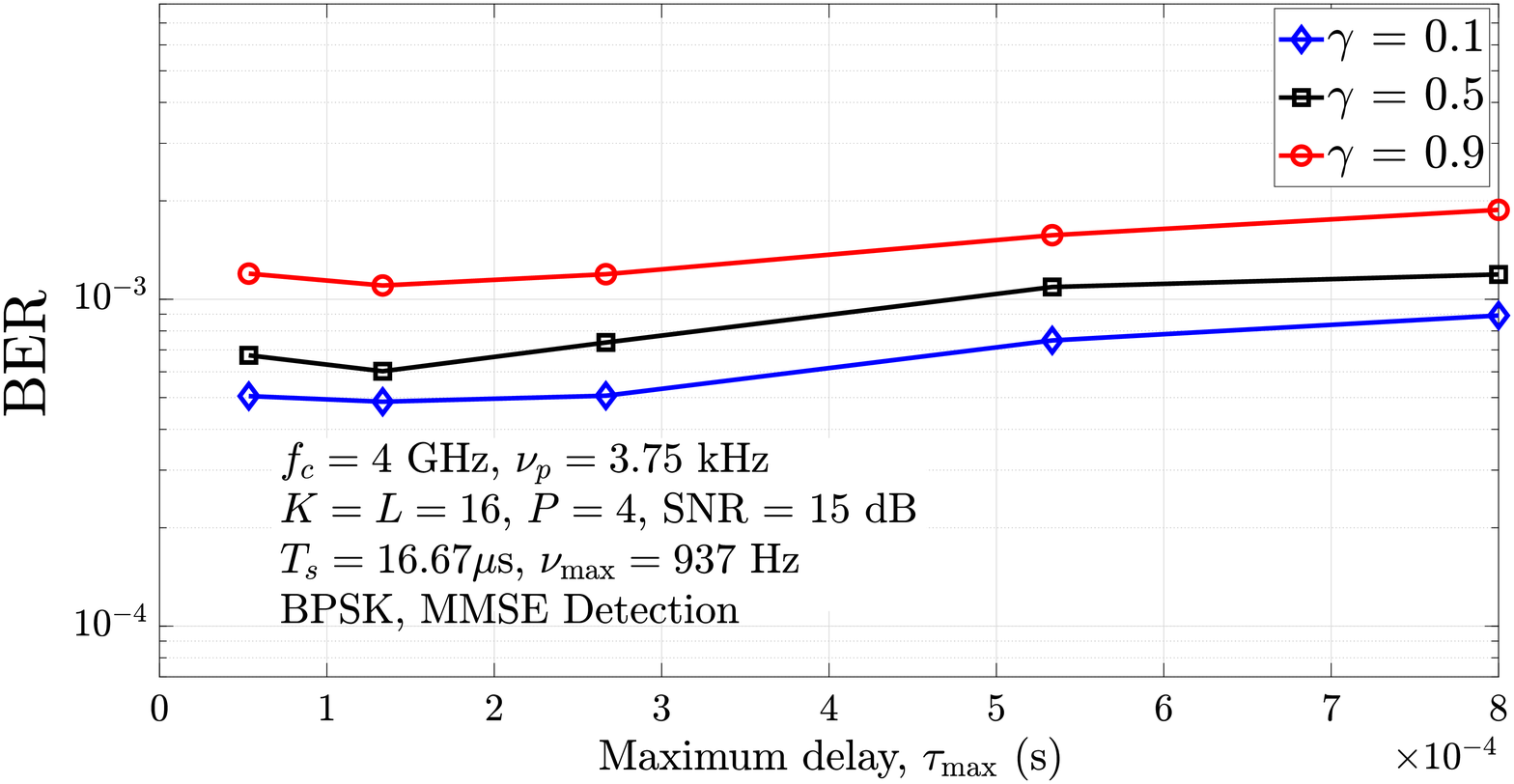}
\vspace{-4mm}
\caption{Effect of RC pulse roll-off factor, $\gamma$, on BER vs $\tau_{\max}$ performance.}
\vspace{-4mm}
\label{DZT_BER_vs_delay}
\end{figure}

The choice of roll-off factor ($\gamma$) of the RC pulse plays a crucial role in the performance of DZT-OTFS. This is due to the DD domain spread of the RC pulse. As $\gamma$ increases from $0$ to $1$, the spread of the RC pulse in the transform domain increases considerably. This spread of the DD domain symbols into the adjacent bins lead to a degraded BER performance. The simulation results in Figs. \ref{DZT_BER_vs_Doppler} and \ref{DZT_BER_vs_delay} illustrate this point, where the BER is shown as a function of $\nu_{\max}$ and $\tau_{\max}$, respectively, for $\gamma=0.1,0.5,0.9$ at an SNR of 
15 dB. The performance is observed to degrade as the $\gamma$ is increased from 0.1 to 0.9, because of the the increased spread of the RC pulse. 

\subsection{Complexity of DZT-OTFS and two-step OTFS}
Here, we present a complexity comparison between DZT-OTFS and two-step OTFS 
in transforming a time domain sequence to DD domain and vice-versa. For DZT-OTFS, from \eqref{DZT_eqn}, we observe that for a given $(k,l)$, $Z_{y}[k,l]$ can be calculated for all $k=0,1,\cdots ,K-1$ using $K$-point discrete Fourier transform (DFT), whose complexity is $O(K\log_2K)$. Therefore, for $k=0,1,\cdots ,K-1$ and $l=0,1,\cdots ,L-1$, the overall complexity for DZT-OTFS is $O(KL\log_2K)$. For two-step OTFS, transforming a signal from DD to TD via TF domain involves SFFT, whose complexity is of the order $O(KL\log_2KL)$, which is higher than that of DZT-OTFS.

 \vspace{-1mm}
\section{Conclusions}
\label{sec5}
\vspace{-1mm}
In this work, we presented an early investigation of the bit error performance of OTFS modulation realized using discrete Zak transform approach. We derived a compact DD domain input-output relation for DZT-OTFS in matrix-vector form, which is valid for fractional delay-Dopplers and practical pulse shapes. This is a new and useful contribution as the derived matrix-vector representation of the end-to-end DD domain input-output relation gives a basic foundation for exploring further into the design of efficient techniques and algorithms for DZT-OTFS transceivers. Our simulation results showed that DZT-OTFS can achieve better performance compared to two-step OTFS over a wide range of Doppler spreads at a lesser complexity. In this work, we presented the diversity order results through explicit computation of the rank profile of the difference matrices. A formal analytical proof on the diversity order results is a topic  for future work. Also, development and performance evaluation of efficient algorithms for DD domain equalization/detection and channel estimation for DZT-OTFS remains open for future research.

\end{document}